# Quaternary crystals CdZnTeSe: Growth via the vertical Bridgman method with different compositions of raw materials

*S.V. Naydenov[*,1], O.K. Kapustnyk[1], I.M. Pritula[1], D.S. Sofronov[1], I.S. Terzin[1], N.O. Kovalenko[1,2]*

[1] Institute for Single Crystals of the National Academy of Sciences of Ukraine,
60 Nauky Ave., 61072 Kharkiv, Ukraine
[2] Helmut-Schmidt-Universität, Holstenhofweg 85, 22043 Hamburg, Germany

**Abstract**

Indium-doped semiconductor crystals CdZnTeSe with several different compositions of raw materials were grown via the vertical Bridgman method under high-pressure argon. For the first time, these crystals were obtained via a combined method from a mixture of simple and binary starting components. A theoretical analysis of the permissible reactions for obtaining multicomponent CdZnTeSe crystals from different compositions of starting materials was performed. The homogeneity of the distribution of the atomic composition and electrical resistance (in the dark and under illumination) of the obtained crystals was studied. Crystals grown via the new combined method presented the best homogeneity of composition and electrophysical properties.

## *1. Introduction*

CdTe-based semiconductor crystals are widely used as infrared and radiation detectors. The ternary crystal CdZnTe (CZT) is well known and has been used for X-ray and gamma-ray detection at room temperature [1]. However, it is among the semiconductor crystals that are difficult to grow [2]. Various drawbacks [3] and the relatively low yield of detector material during CZT growth are related to the "poor" thermophysical properties of the melt-crystal system, as well as the "soft-brittle" structure of the mother CdTe matrix. During the growth of these crystals, there is a greater tendency for the formation of intrinsic and impurity point defects, especially at elevated growth temperatures and/or melt contamination. They have low activation energies of extended (growth) defects, a large degree of ionic chemical bonding, strong volatility of the Fermi level in the forbidden zone when the impurity composition changes, and non-stoichiometry of the composition owing to strong segregation and/or increased volatility of some atomic components.

Long-term efforts aimed at improving CZT crystal fabrication technology have not led to the expected results, although these crystals continue to be widely grown, mainly from solution melts, via the THM method. Recently, new quaternary CdZnTeSe (CZTS) crystals have been proposed as alternatives [4]. Theory [5] and experiments [6, 7] indicate that there is a "hardening" effect due to the addition of a small amount of selenium. The thermodynamic lowering of the excess Gibbs energy of solid solution formation when selenium is added makes it less energetically advantageous to disrupt the crystal structure. As a result, the number of the most harmful defects (clusters of small grains, dislocation networks, large tellurium inclusions, etc.) may decrease in the CZTS crystals.

Although CZTS crystal growth methods are mainly (apart from some technological nuances) the same as CZT crystal growth methods are, there are some differences. In melt methods such as High-Pressure Bridgman Method (HPBM) [8-11] or Vertical Freezing Gradient (VFG) method [12], ternary CZT crystals are usually synthesized and/or grown from a mixture of three simple components: cadmium, tellurium and zinc. In this case, the large zinc segregation coefficient 1.24-1.35 (see, for example, [2]) leads to significant inhomogeneity in the ingot composition. The synthesis of CZT by the addition (in loading) of zinc to the binary compound CdTe gives worse results. In the Travelling Heater Method (THM) [13, 14], CZT is grown by zone melting from solution-melt of excess tellurium (more than 50% of the total solution) containing dissolved cadmium and zinc. As a growth raw material, either simple substances or a charge of presynthesized CZT polycrystals are preferably used herein. CZTS quaternary crystals contain four atomic elements: cadmium and zinc, tellurium and selenium. This increases the allowable set of combinations of starting raw materials. Potentially, such crystals can be grown from significantly different compositions of starting materials (see below for more details). Ternary CZT crystals can be grown from both uncompounded starting materials (a mixture of cadmium, zinc and tellurium) and compounded starting materials (e.g., a mixture of cadmium telluride with zinc or the CZT substance itself). In the first case, the melting temperature of the individual components of the starting raw material is different; in the second case, it can be the same if a single crystal is grown from a polycrystalline material. CZTS crystals are more often grown from a complex composition of several starting components. The melting temperatures and mutual solubilities of these materials in the melt can vary greatly. This feature initially makes it difficult to obtain a homogeneous crystalline material. However, it is from homogeneous crystals that the best quality detector elements are made.

[*] sergei.naydenov@gmail.com

The specificity of the growth of multicomponent semiconductor A2B6 crystals is that they are highly sensitive to changes in growth conditions, including the choice of loading of initial growth components. Typically, CZTS crystals are grown from starting material in the form of a composition of "CZT+CdSe" ternary CZT with a selenium precursor [15] or a composition of four simple components, "Cd+Zn+Te+Se", e.g., [16], or a composition of binary components, "CdTe+ZnTe+CdSe", e.g., [17]. The question naturally arises as to which of the methods of starting material selection gives the best results in terms of obtaining the most homogeneous crystalline ingots, with a high yield of useful material and acceptable properties. Previously, a similar problem was considered for CZT ternary crystals [18]. In this work, quaternary CZTS crystals with different compositions of growth components were grown via the vertical Bridgman method under high argon pressure. A comparison of their homogeneity with respect to compositional changes and electrophysical properties was performed. This allows us to identify two promising ways of growing CZTS crystals from the melt, with the starting material consisting only of binary components CdTe, ZnTe, and CdSe and a new *combined way of* growing from the composition of two simple components, Cd and Te, and two binary components, ZnTe and CdSe. Note that the combined method, to the best of our knowledge, has not been previously proposed.

## 2. *Theoretical analysis*

The same CZTS solid solution as a chemical compound can be obtained via different reactions. For example, the reaction of synthesizing a solid solution from its constituent binary components can be used

$$(1-x-y)CdTe + xZnTe + yCdSe \rightarrow Cd_{1-x}Zn_xTe_{1-y}Se_y, \qquad (1)$$

or synthesis from simple components alone

$$(1-x)Cd + xZn + (1-y)Te + ySe \rightarrow Cd_{1-x}Zn_xTe_{1-y}Se_y, \qquad (2)$$

or synthesize a ready-made CZT ternary compound with a selenium precursor in the form of CdSe

$$(1-y)Cd_{1-x}Zn_xTe + yCdSe \rightarrow Cd_{1-x+xy}Zn_{x-xy}Te_{1-y}Se_y. \qquad (3)$$

In the latter case, $Cd_{1-x+xy}Zn_{x-xy}Te_{1-y}Se_y \approx Cd_{1-x}Zn_xTe_{1-y}Se_y$ provided that the atomic concentrations of $x \ll 1$ and $y \ll 1$ of the solid solution are small, when their product can be neglected as a value of a higher order of smallness, $xy \ll \{x, y\}$. Other reactions are also possible.

The most general type of reaction for the synthesis of CZTS from binary components is as follows:

$$(1-x-y-z)CdTe + xZnTe + yCdSe + zZnSe \rightarrow Cd_{1-x-z}Zn_{x+z}Te_{1-y-z}Se_{y+z}. \qquad (4)$$

The synthesis of a material from several simple and binary components is possible. For example, one can formally use such a reaction

$$(1-x-y)Cd + (1-x-y)Te + xZnTe + yCdSe \rightarrow Cd_{1-x}Zn_xTe_{1-y}Se_y. \tag{5}$$

Even more exotic is the reaction that is obtained by combining the reactions (1) and (5). The reagents supplying cadmium and tellurium atoms to the system include both the simple components themselves and their binary component CdTe:

$$(1-x-y-z)CdTe + xZnTe + yCdSe + zCd + zTe \rightarrow Cd_{1-x}Zn_xTe_{1-y}Se_y. \tag{6}$$

Note that here, the concentration $z$ of additional simple components remains arbitrary. The full list of permissible reactions is rather wide and is not given here. Despite this diversity, practical conditions for CZTS crystal growth impose certain restrictions on the choice of initial reagents.

Considering the general reaction (4), the binary compound ZnSe could be used to introduce zinc and selenium into CZTS. However, ZnSe has a high melting point (1525°C), which far exceeds the melting point of the other growth components and the melting point of CZTS (1150-1180°C) at which crystals grow from the melt. In addition, almost all the chalcogenides under consideration are very poorly soluble even at high temperatures. In addition to purely molecular reasons, this is due to the high viscosity of melts based on cadmium selenide. Moreover, high-melting chalcogenides dissolve much worse than the other chalcogenides do. For these reasons, the growth of monophase CZTS crystals via the melt method becomes impossible if the binary component ZnSe is present in the composition of the raw material even in insignificant amounts (a few atomic or mass percentages). We confirmed this when trying to grow such CZTS crystals via the Bridgman method.

It is possible to incorporate selenium into CZTS crystals directly via the reaction (2). However, the simple components zinc and selenium are present in the raw material. At high melt temperatures, the exothermic reaction of zinc selenide formation can start. It is accompanied by the release of large amounts of heat and is explosive in nature. Therefore, for a more controlled introduction of selenium into the CZTS, we chose the method of adding the binary component CdSe. The disadvantage of this method is the excessive contamination of cadmium selenide [19], for which, even with modern purification methods, the required purity of 99.9999% (6N), which is inherent to all other growth components, has not yet been achieved.

The growth of CZTS crystals from the CZT compound via the Bridgman method (3) has not been shown to produce crystalline ingots with a uniform distribution of selenium. The selenium concentration can vary several times along the ingot (see Section 4). In our opinion, this is also due to the poor mutual solubility of chalcogenides in the melt. Moreover, this negative effect increases in the transition from binary to ternary compounds, i.e., from the CdTe melt to the CZT melt.

Finally, CZTS can be grown (synthesized) via a complex reaction of type (5), in which both simple and binary components participate simultaneously. Strictly speaking, in this method of growth, CdTe is first synthesized from the melt of cadmium and tellurium (elementary bimolecular chemical reaction), and only then, at sufficiently high temperatures and after complete dissolution of all binary components ZnTe and CdSe in the CdTe melt can CZTS crystals grow. The advantage here is that the binary components are much better dissolved in cadmium, and tellurium melts at intermediate synthesis stages. This generally leads to better mutual dissolution of all the components and homogenization of the final melt.

It is useful to compare the thermodynamic equilibrium constants for the reactions (1) and (5). For simplicity, let us consider the CZTS melt as an ideal solution. Then, for saturated vapors above the melt, Raoul's law [20] is satisfied, and their pressure is proportional to the concentration of components in the melt. This allows us to write the equilibrium constant for the reaction of CZTS synthesis from the composition of binary components in the form of

$$K_{binary}(T,p) = \left[\frac{p_{CZTS}}{p^0_{CZTS}}\right] \left\{ \left[\frac{p_{CdTe}}{p^0_{CdTe}}\right] \left[\frac{p_{ZnTe}}{p^0_{ZnTe}}\right] \left[\frac{p_{CdSe}}{p^0_{CdSe}}\right] \right\}^{-1}, \tag{7}$$

and for the combined synthesis in the form

$$K_{comb}(T,p) = \left[\frac{p_{CZTS}}{p^0_{CZTS}}\right] \left\{ \left[\frac{p_{Cd}}{p^0_{Cd}}\right] \left[\frac{p_{Te_2}}{p^0_{Te_2}}\right]^{1/2} \left[\frac{p_{ZnTe}}{p^0_{ZnTe}}\right] \left[\frac{p_{CdSe}}{p^0_{CdSe}}\right] \right\}^{-1}, \tag{8}$$

where $T$ is the temperature, $p$ is the external pressure, $[p]$ is the saturated vapor pressure of partial components above the melt, and the upper index "0" denotes bringing the system to standard conditions. When obtaining the formula, the fact was taken into account that when evaporating from the melt, atomic tellurium is transformed into a molecular diatomic gas. Combining these expressions, we obtain the relation

$$K_{binary}(T,p) \approx K_{comb}(T,p) \left[\frac{p_{Cd}}{p_{Cd}^0}\right] \left[\frac{p_{Te_2}}{p_{Te_2}^0}\right]^{1/2} \left[\frac{p_{CdTe}}{p_{CdTe}^0}\right]^{-1}. \qquad (9)$$

At the CZTS melting temperature, the vapor pressure of cadmium above the melt reaches several tens of atmospheres (the most volatile component), the vapor pressure of tellurium reaches several atmospheres, and the vapor pressure of cadmium telluride does not exceed the atmospheric pressure (approximately $10^5$ Pa). Therefore, the condition is satisfied with good accuracy:

$$K_{comb}(T,p) \sim 10^{-2} K_{binary}(T,p) \ll K_{binary}(T,p). \qquad (10)$$

In the case of a weakly concentrated CZTS solid solution, $x \ll 1, y \ll 1$, the Gibbs energies for the reactions under consideration can be related to the simple expression

$$\Delta G_{comb} \approx \Delta G_{binary} + RT \ln\left(\frac{Q_{comb}}{Q_{binary}}\right), \qquad (11)$$

where $\Delta G$ is the difference in the chemical potentials of the products and reactants taking into account stoichiometric coefficients (it is assumed that the reaction proceeds, $\Delta G \leq 0$); $Q$ is the quotient of the corresponding reactions; $R$ is the gas constant. When approaching a state of thermodynamic (chemical) equilibrium, the factor $Q$ tends to the equilibrium constant. For simplicity, let us assume that $Q \approx K(T,p)$. Then, the formulas (10) and (11) follow the relation

$$|\Delta G_{comb}| = |\Delta G_{binary}| + \alpha RT, \qquad (12)$$

where $\alpha \sim 1$ is on the order of unity. The difference in the Gibbs energies of solid solution formation for these reactions can reach values of several units $RT$, which are quite large at characteristic CZTS melt temperatures. The ratio (12) indicates that the Gibbs energy of the combined reaction decreases much more strongly when the parameters of the system change (approaching equilibrium) than when the reaction is based on binary components. At the physical level, this corresponds to faster synthesis and uniform homogenization of the CZTS melt under otherwise equal growth conditions. As a result, the ability to obtain more homogeneous CZTS crystals is promoted.

Considering the results of the theoretical analysis, among all possible methods of CZTS crystal production from melt, the two most promising variants should be identified: *synthesis from binary components* by the reaction (1) and *combined synthesis* by the reaction (5). The experimental growth of the CZTS crystals confirmed this conclusion. Both the synthesis and the final growth of CZTS crystals from the melt are carried out in the same growth process to avoid contamination when the synthesized material is reloaded and to reduce material and time costs.

### *3. Experimental part*

CZTS crystals with diameters of up to Ø25 mm (1 inch) and lengths of up to 60 mm were grown via the vertical Bridgman method under high-pressure argon (6N). Crystals were grown in a single-zone furnace with automatic control of the temperature and crucible speed. The temperature in the melt was monitored by a W-Re thermocouple. Growth raw materials of 99.9999% (6N) purity were used, except for a CdSe charge of 99.9995% (5N5) purity. The raw materials were not pulverized before loading. The phase composition for different compositions of the starting raw materials was calculated from the stoichiometric formula of the solid solution $Cd_{0.9}Zn_{0.1}Te_{0.98}Se_{0.02}$ and corresponded to the same concentration of zinc and selenium in the loading for all the growths.

The starting material was loaded under high-purity conditions into a precleaned graphite crucible made of dense low-ash graphite of 5N-6N purity. On several occasions, including the growth of CZTS from presynthesized CZT, a quartz ampoule with an inner diameter of Ø25 mm was used as a crucible. The optimal growth regime was selected on the basis of the choice of starting material and the use of different synthesis reactions. The melt temperature in all the cases was slightly above the melting point of CZTS, and the temperature gradient at the crystallization front was chosen to be 20°C/cm. Prior to the crystal growth stage, the melt was incubated for 24 h for maximum homogenization. The argon pressure in the active phase of growth reached 22–24 atm. No seed crystals were used. Spontaneous nucleation starts from the conical part of the crucible (spout) and then continues on its sidewalls. The crystal growth rate did not exceed 0.8 mm/hour. At the optimum choice of mode at an output, a large-block crystalline ingot with small twinning in monocrystalline grains was obtained. Growth losses due to material entrainment usually do not exceed 3–6%. When the crystals were grown in quartz ampoules, material entrainment was much greater than when they were grown in graphite crucibles. To compensate for doping, we used an indium donor impurity with a targeted concentration in the crystal of approximately 10 ppm on average. This allows us to obtain high-resistivity indium-doped CZTS:In crystals.

The used growth mode was the same for all crystals and mainly did not depend on the choice of starting material. Figures 1-4 show photographs of the obtained CZTS:In crystal ingots, 5 mm thick longitudinal plates were cut from them. The plates were cut from the central part of the ingots. After the plate was cleaned from impurities, the crystalline quality of the material was visually evaluated, including the composition of single-crystalline bars, the presence of twins, clusters of small grains, growth inclusions and seizures, and other visually visible defects. X-ray powder diffractometry of the obtained solid solution crystals confirmed their monophase composition. X-ray studies of the crystal structure of the obtained crystals (rocking curves, etc.) will be carried out in further studies.

Mapping of the atomic composition and electrical resistivity in the dark and under illumination was performed for mechanically processed plates (chemical polishing was not applied). The spatial step of the measurements along the ingot axis was 4–5 mm. The atomic composition variations were determined via electron density X-ray spectroscopy on a low-vacuum scanning electron microscope (JSM-6390LV) with an AZtech Energy X-max 50 detector analyzer. The accuracy of the composition measurements is estimated to be approximately 0.5 at.%. When the composition was mapped, areas containing tellurium phases or other micro-inclusions not characteristic of the matrix itself were excluded from the field of observation.

The variation in the electrophysical parameters was investigated via a noncontact expression method (without applying ohmic contacts to the surface of the samples) on a special stand with an E6-13A tera-ohmmeter (USSR) at a constant applied voltage of 100 V. The direct current was applied step by step from opposite sides of the plate at points opposite each other. Special current collectors with clamping silicone rings were used for contact to protect against possible surface leakage currents. The stand was recalibrated on test samples of high-resistance semiconductor crystals with known resistance. The measurements at each step were repeated several times to improve the statistics. The errors in measuring the resistance of high-resistivity samples ($10^{11}$-$10^{12}$ Ohm) are estimated to be on the order of giga-Ohm. The conductivity type was controlled by measuring the thermal EMF (the Seebeck effect). All the samples were found to have p-type conductivity.

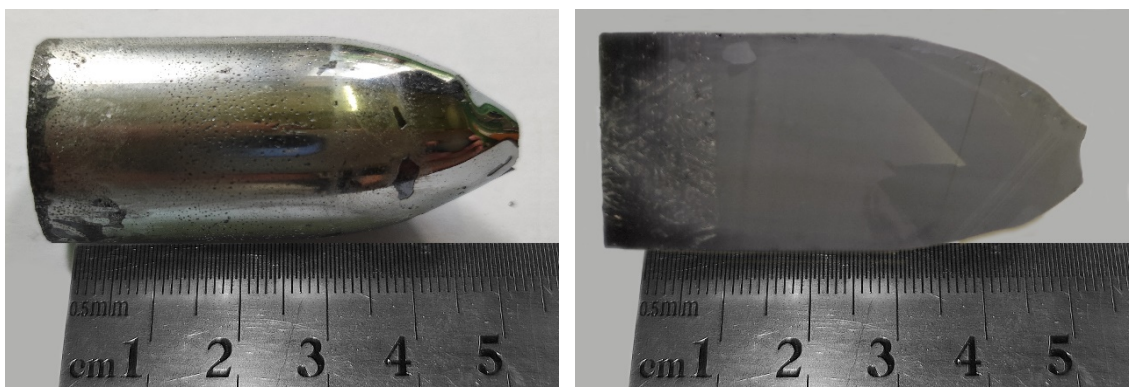

Fig. 1. CZTS-2 crystal ingot (a) and cut plate (b) with the starting composition "CZT:In + CdSe".

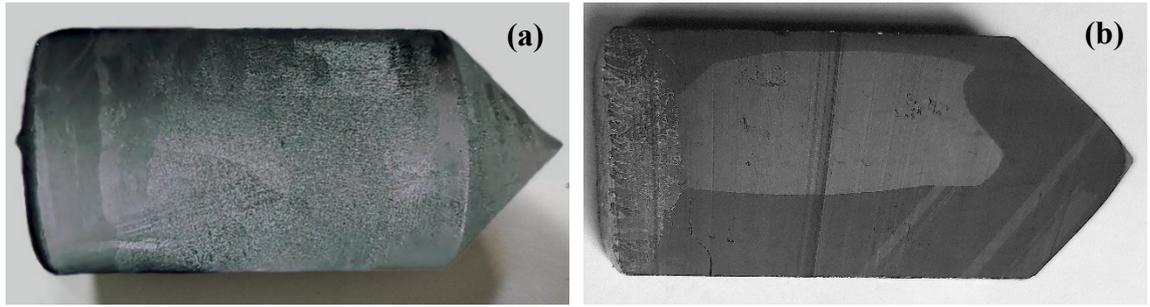

Fig. 2. CZTS-5 crystal ingot (a) and cut plate (b) with the starting composition "CdTe+ZnTe+CdSe+In".

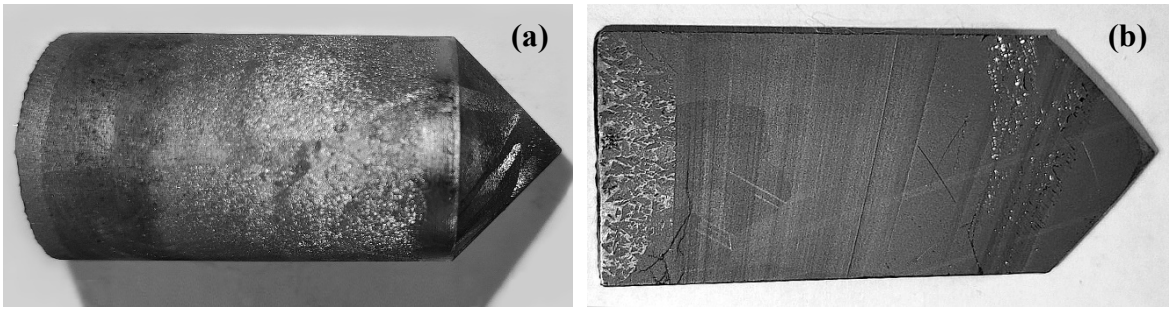

Fig. 3. CZTS-7 crystal ingot (a) and cut plate (b) with the starting composition "CdTe+ZnTe+CdSe+Cd+Te+In".

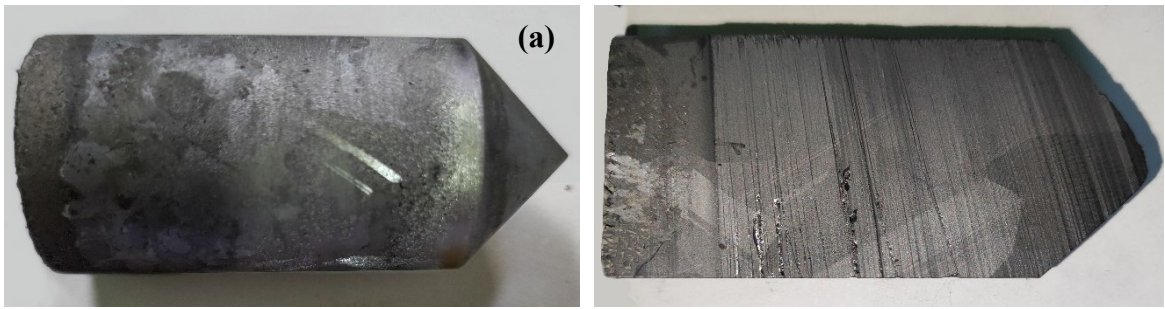

Fig. 4. CZTS-8 crystal ingot (a) and cut plate (b) with the starting composition "Cd+Te+ZnTe+CdSe+In".

## *4. Results and discussion*

The results of atomic composition mapping of CZTS crystals (see Figs. 1-4) obtained from different starting materials are presented in Fig. 5. The diagrams indicate the component contents of zinc and selenium in atomic percentages. The distances along the abscissa axis correspond to the measurement points at the transition point along the ingot axis from the spout to the heel. Variations in the cadmium and tellurium components along the ingot were obtained but are not given here. The relative values are not as significant as the variations in the zinc and selenium contents are, except for the transition region 10-15 mm long at the end of the ingot. The boundary of this region is clearly visible in Figs. 1-4. In this growth zone, the melt solidification is too fast, the stoichiometry is strongly disturbed (a large excess of tellurium is accumulated), and no crystals are formed. As the measurements show, the total molar concentrations of the cations "Cd+Zn" and anions "Te+Se" are always close to unity. Deviations in the cation-anion balance usually do not exceed 1 at.% and are explained by the total accuracy of the measurements – 0.5 at.% each of the errors in determining the composition of cations and anions.

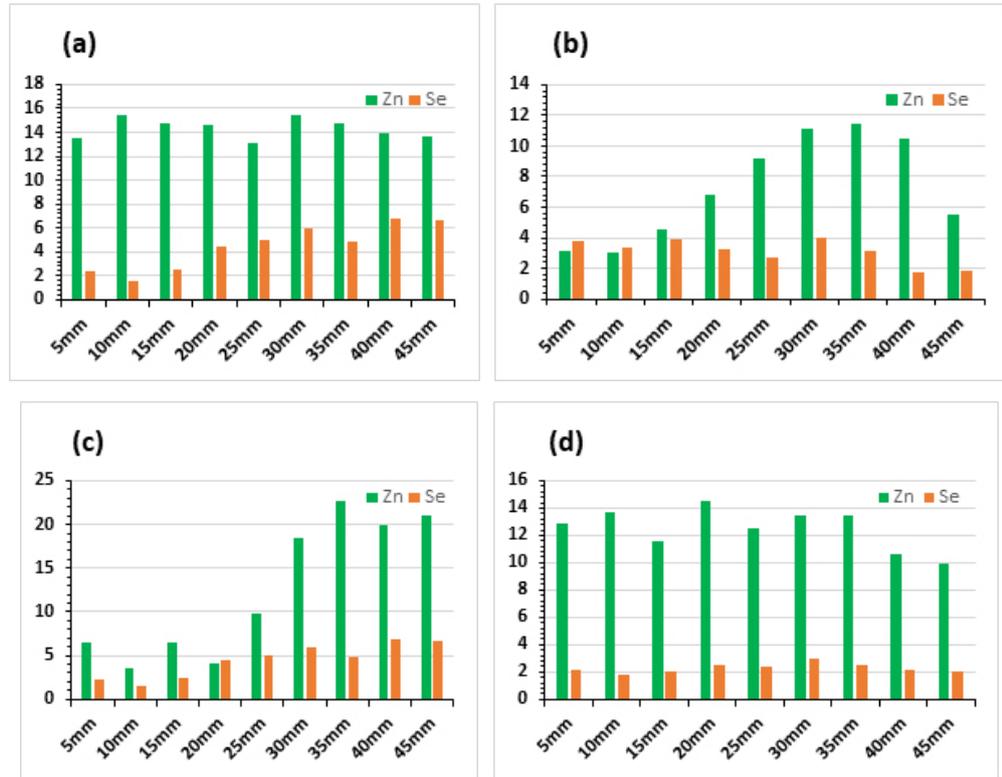

Fig. 5. Spatial distributions of Zn (at.%) and Se (at.%) across the grown crystal ingots: CZTS-2 (a); CZTS-5 (b); CZTS-7 (c); CZTS-8 (d).

As shown in Fig. 5, the selenium distribution is insufficiently homogeneous for the CZTS-2 crystal grown from the starting composition "CZT+CdSe" and the CZTS-7 crystal grown from the composition "CdTe+ZnTe+CdSe+Cd+Te". For the CZTS-5 crystal grown from the composition "CdTe+ZnTe+CdSe+CdSe", the selenium distribution is more homogeneous. The CZTS-8 crystal grown from the "Cd+Te+ZnTe+CdSe" composition is almost ideal.

The distributions of zinc in the CZTS-5 and CZTS-7 crystals are highly inhomogeneous. This is a consequence not only of the large segregation of zinc at the crystallization front but also, perhaps to an even greater extent, of the poor solubility of binary components in the cadmium telluride melt. Increasing the time of melt homogenization to overcome these harmful effects is not advisable because it leads to strong evaporation of volatile components (cadmium and zinc) and a violation of stoichiometry. The distribution of zinc in the CZTS-2 crystal is more homogeneous. This is due to the growth of the CZTS crystals from the selected pre-synthesized polycrystalline CZT, in which the zinc distribution was initially quite homogeneous. Notably, the CZTS-8 crystals obtained via the combined method without pre-synthesis have high homogeneity in terms of the zinc component distribution. A slight decrease in the zinc content here occurs only at the end of the crystal and is due to the segregation effect.

In comparison with the results of other works on growth of homogeneous CZTS crystals using the Bridgman method, we have obtained more homogeneous crystals. In our crystals, the relative variation of zinc content (the element with the highest segregation coefficient) along the ingot is within 5-15% of the nominal value, whereas when grown using other methods, this variation often exceeds 20-30% (see, for example, [23]), and sometimes can even approach up to 50-60% (see, for example, [24]).

Thus, a comparison of the compositions of the obtained crystals indicates that the CZTS crystals obtained via the combined method from the starting material "CdTe+ZnTe+CdSe+Cd+Te" have the highest homogeneity in terms of the Zn and Se components. The physical reasons for the appearance of such homogeneity were explained on the basis of thermodynamic reasoning at the end of Section 2. These results are related to the higher rate of decrease in Gibbs energy during the formation of a solid solution from the indicated composition of reagents (starting raw materials).

Measurements of the composition of the grown CZTS:In crystals were complemented by measurements of their electrical resistivity in the dark and in the light under illumination with a 100 W halogen lamp. These data are shown in Fig. 6. Resistivity mapping of the plates cut from the crystals was performed at the same points as the atomic composition measurements. The concentration of the indium doping additive (at raw material loading) for these crystals differed in several times. Considering the very low distribution coefficient of In (not more than 0.1), its concentration was approximately an order of magnitude greater in the loading state (30 to 100 ppm) than in the crystal state.

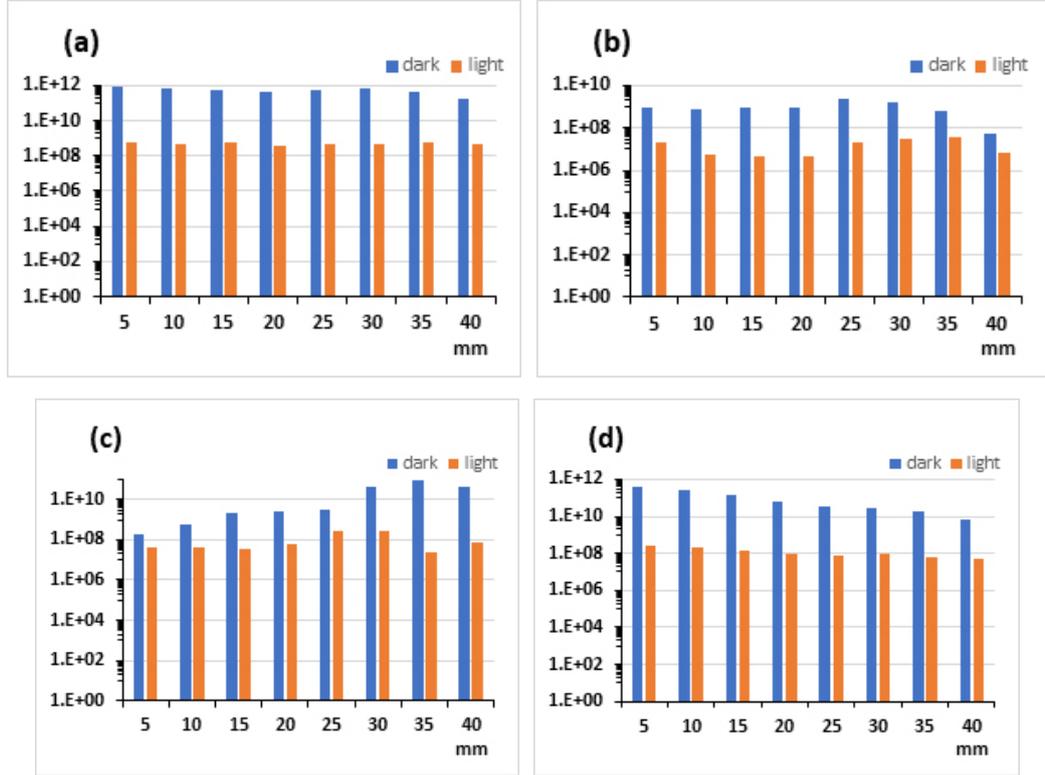

Fig. 6. Spatial distributions of electric resistance (Ohm) in the dark and photo-resistance under illumination across the grown crystal ingots: CZTS-2 (a); CZTS-5 (b); CZTS-7 (c); CZTS-8 (d).

The variation in the specific electrical resistance of a material depends on many factors. The variation in the dark resistance mainly depends on the position of the Fermi level in the forbidden zone. If full compensation has not occurred during doping of the crystal, i.e., the Fermi level is still far from the middle of the forbidden zone, the change in composition does not greatly affect the resistance variation. These low-resistivity CZTS samples have low resistivities (typically $10^6$–$10^8$ Ohm·cm or lower) and are virtually indistinguishable from the resistivity under illumination. If the compensation is strong enough, the dark resistivity is sensitive to variations in the crystal composition, especially to variations in the zinc content of the crystal, since the forbidden band width changes [21]. However, the strongest variations in the dark resistivity are related to variations in the concentration of the doped (compensating) impurity.

Segregation of the impurity at the crystallization front leads to its inhomogeneous distribution in the crystal. With good melt mixing, the impurity distribution in the crystal is described by the Sheil formula [22]:

$$C(x) = kC_0 (1-x)^{k-1}, \qquad (13)$$

where $C$ is the concentration of impurity in the crystal, $C_0$ is the initial concentration of impurity in the load, $k$ is the distribution (segregation) coefficient, and $x$ is the fraction of volume that has crystallized.

In this case, the parameter $x$ corresponds with good accuracy to the longitudinal coordinate along the crystal growth axis, i.e., at the points of composition and resistivity measurements. Considering the smallness of the indium distribution coefficient $k \ll 1$, the expression (13) can be simplified as follows:

$$C(x) \approx kC_0(1+\alpha x), \qquad (14)$$

where coefficient $\alpha$ is a correction factor associated with deviations of the real impurity distribution from the model formula (13), including those due to inhomogeneous initial impurity loading and poor melt mixing. Usually, to avoid strong inhomogeneity (in the crystal), the indium in the loading mixture is placed entirely in the lower part of the crucible, from which crystal growth begins. During crystal growth, the excess indium, which constantly arises at the crystallization front, easily diffuses into the initially indium-depleted region of the melt. As a result, there is an effective equalization of the indium concentration and partial compensation of the segregation effect. The indium content in the obtained crystal often decreases from the beginning to the end of the ingot. In contrast, with a uniform initial distribution of indium, the indium concentration should increase from the beginning to the end of the ingot.

The resistivity of a semiconductor depends exponentially on the position of the Fermi level relative to the edges of the forbidden zone. In a compensated (high resistivity) semiconductor, the Fermi level is near the middle of the forbidden zone. With a small change in the concentration of the doped (compensating) impurity, the Fermi level shifts in the forbidden zone in proportion to this concentration. As a result, the resistivity changes in accordance with the local change in the concentration of the doped impurity. Moreover, the logarithm of the dark resistivity in the considered case should change in a linear way, similar to the change in the concentration of the doping impurity in the crystal:

$$\ln \rho(x) \approx (1+\beta x)\ln \rho_0, \qquad (15)$$

where $\beta$ and $\rho_0$ are constant values for this crystal.

Fig. 7 shows the distributions of the logarithm of resistance for the CZTS crystals. Since the test samples were fabricated as identical plates, their resistivity when measured at different points is proportional to the normal resistivity with good accuracy (we neglect leakage currents). The dark resistivity of the CZTS-2 sample does not vary much along the ingot. This is due to the elevated (100 ppm in the initial CZT synthesis feed) and homogeneous indium content in the starting material from which CZTS-2 was grown. Here, there is strong compensation for the position of the Fermi level in the forbidden band. The resistivity in the ingot becomes maximal, and its variations are hardly noticeable. This corresponds to the case where $\alpha \ll 1$ in the formula (14). In contrast to the CZTS-2 sample, the CZTS-5, CZTS-7, and CZTS-8 crystals had the same moderate loading of 40 ppm indium during growth. The material compensation was incomplete. As a result, the CZTS-5 and CZTS-7 crystals exhibit strong heterogeneity in the distribution of (logarithm of) resistivity across the ingot, which does not agree well with the expected relationship. Particularly strong variations are inherent in the CZTS-7 crystal. This is due to poor melt homogenization, which leads to significant inhomogeneities in both the basic composition and the indium impurity in the crystal ingot. At the same time, the CZTS-8 crystal shows an almost linear dependence on the type , which corresponds to the formula (14). Consequently, it is the most "homogeneous" from the point of view of variation in electrophysical properties. Strictly speaking, with respect to electrical resistivity, the CZTS-8 crystal has high homogeneity (local homogeneity in some parts of the ingot) but low uniformity (global homogeneity throughout the ingot). Its global heterogeneity is due to the strong segregation of indium, which was insufficient for loading; therefore, the compensation of the Fermi level was incomplete in the whole crystal. By increasing the concentration of indium in the loading mixture to 60-100 ppm, it is possible to reduce this harmful effect.

In contrast to dark resistance, the photoresistance of crystals under a sufficiently powerful irradiation source depends to a greater extent not on the concentration of fully ionized impurities but on the concentration of generated intrinsic charge carriers (electrons and holes). Their concentration at the same illumination intensity depends on the width of the forbidden zone. Therefore, the crystals that are the most homogeneous in composition (whose forbidden band width is almost unchanged) have the smallest variations in photoresistance. A precise comparison of the data in Fig. 6 reveals that the average relative variations in the resistivity of the CZTS-5, CZTS-7, and CZTS-8 crystals are 74%, 82%, and 34%, respectively. Thus, the CZTS-8 crystal has the best resistance homogeneity both in the dark and in the light.

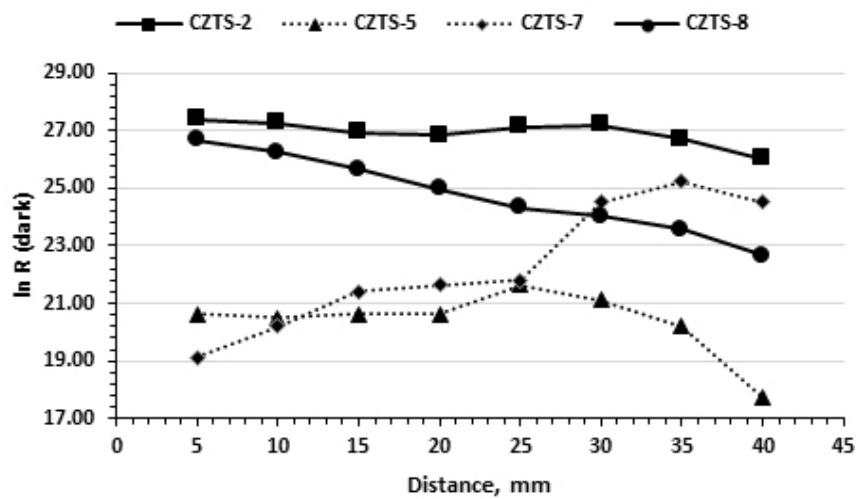

Fig. 7. Spatial dependence of the logarithm of the dark resistance (Ohm) for different crystal ingots.

Taking into account the obtained results, we can conclude that the crystal grown by the combined method from the starting material "Cd+Te+ZnTe+CdSe" has the best homogeneity of electrophysical parameters.

## 5. Conclusions

The study of CZTS:In quaternary crystals grown via the melt method from different compositions of growth raw materials have shown that the crystals with the best homogeneity in composition (almost constant distribution of zinc and selenium component concentrations in the crystalline ingot) and highly homogeneous electrophysical properties (proportionality of the logarithm of the dark resistivity of the material to the concentration of compensating indium impurity and low variations in photoresistivity in different parts of the crystalline ingot) are obtained by the growth from the combined starting material "Cd+Te+ZnTe+CdSe". This is due to better dissolution of binary components in the mixed melt of cadmium and tellurium, as well as better homogenization of the resulting CZTS melt under otherwise identical growth conditions. The physical mechanism of these effects is due to the higher rate of decrease in the Gibbs energy of solid solution formation when the system strives for a state of thermodynamic equilibrium in comparing to other reactions allowed for the synthesis of multicomponent crystals. Quaternary CZTS crystals with satisfactory homogeneity are also obtained by growing from the starting composition of binary components "CdTe+ZnTe+CdSe". However, to eliminate significant compositional variations inherent in such crystals, it is necessary to use special growth techniques, which are discussed separately.

*Acknowledgements*

We kindly acknowledge to Dr. Pavel V. Mateichenko for performing the EDS-measurements by the scanning electron microscope.